\documentclass[11pt,a4paper,english,superscriptaddress,aps,prd,nofootnoteinthebib,preprint]{revtex4}
\usepackage{amssymb}
\usepackage{amsmath} 
\usepackage{slashed}
\usepackage{graphicx}
\usepackage{babel}
\usepackage{hyperref}
\usepackage{color}
\usepackage{comment}
\newcommand{\gdualn}[1]{\overset{\:{}^{{}^{\boldsymbol{\neg}}}}{\smash[t]{#1}}} 
\newcommand{\0}{\mbox{\boldmath$\displaystyle{\bf O}$}}
\makeatletter\AtBeginDocument{\let\@elt\relax}\makeatother

\begin{document}

\title{Perturbative aspects of mass dimension one fermions non-minimally coupled to electromagnetic field}

\author{Willian Carvalho}
\email{willian.carvalho@icen.ufpa.br}
\affiliation{Faculdade de F\'isica, Universidade Federal do Par\'a, 66075-110, Bel\'em, Par\'a, Brazil.}

\author{M. Dias}
\email{marco.dias@unifesp.br}
\affiliation{Universidade Federal de S\~ao Paulo,
Departamento de F\'isica, Rua S\~ao Nicolau 210,
09913-030, Diadema, S\~ao Paulo, Brazil.}

\author{A.~C.~Lehum}
\email{lehum@ufpa.br}
\affiliation{Faculdade de F\'isica, Universidade Federal do Par\'a, 66075-110, Bel\'em, Par\'a, Brazil.}

\author{J. M. Hoff da Silva}
\email{julio.hoff@unesp.br}
\affiliation{Departamento de F\'isica, Universidade Estadual Paulista, UNESP, Av. Dr. Ariberto Pereira da Cunha, 333, Guaratinguet\'a, SP, Brazil.}

\begin{abstract}
This paper addresses perturbative aspects of the renormalization of a fermion with mass dimension one non-minimally coupled to the electromagnetic field. Specifically, we calculate the one-loop corrections to the propagators and vertex functions of the model and determine the one-loop beta function of the non-minimal electromagnetic coupling. Additionally, we perform calculations of the two-loop corrections to the gauge field propagator, demonstrating that it remains massless and transverse up to this order. We also find that the non-minimal electromagnetic coupling can exhibit asymptotic freedom if a certain condition is satisfied. As a potential dark matter candidate, these findings suggest that the field may decouple at high energies. This aspect holds significance for calculating the relic abundance and freeze-out temperature of the field, particularly in relation to processes involving the ordinary particles of the Standard Model.

\end{abstract}


\maketitle

\section{Introduction}

It dates from over two decades ago, the first version of fermionic mass dimension one fields as candidates to dark matter \cite{Ahluwalia:2004ab}. Since then, the field has undergone modifications in the formulation to conciliate it with Lorentz symmetries (for a broad discussion and physical consequences, see \cite{physrev}). 

Recent advancements in the theory of fermionic fields characterized by mass dimension one have been made \cite{Ahluwalia:2022yvk}. These developments have revealed that the field possesses a two-fold Wigner degeneracy \cite{wig}, effectively doubling its degrees of freedom. Consequently, the field exhibits complete Poincar\`e symmetry. The resulting construction provides a first-principle candidate for dark matter, considering that constraints significantly impact the feasible interactions with standard matter fields, necessitating perturbative renormalizability. While certain potential couplings can still be achieved through a Higgs portal (refer to \cite{Alves:2014kta} for an analysis based on the earlier version of the field), it is conceivable that an additional field associated with a hidden dark sector's $U(1)$ gauge symmetry may exist \cite{dark_universe}.

The renormalization of the previous version of the model, specifically in the absence of gauge interactions, was examined in Ref. \cite{deBrito:2019hih}. The findings from that investigation indicate that the obtained physical results also apply to the current scenario. Building upon this prior work, the objective of our study is to extend the analysis by incorporating a gauge interaction. To accomplish this, we employ dimensional regularization to evaluate the one-loop renormalization of fermions with mass dimension one that are non-minimally coupled to the electromagnetic field. Notably, we consider the presence of a renormalizable non-minimal coupling term $\tilde{e}\gdualn{\lambda} [\gamma_{\mu},\gamma_\nu]\lambda F^{\mu\nu}$, which is allowed by gauge symmetry and has been proposed as a potential source for an effective mass term for the photon \cite{Ahluwalia:2004ab}. The primary result presented in this paper is the demonstration that photon propagation remains massless and transverse up to the two-loop order. This achievement corroborates the transverse aspect of photon self-energy tensor found very recently, via symmetry arguments, in the comprehensive study of Ref. \cite{GNR}. 
 
The structure of this paper is as follows: In the subsequent section, we begin by outlining the fundamental characteristics of the mass dimension one field with Wigner degeneracy and establish the framework of the model. Section III investigates the one-loop self-energy of the mass dimension one fermion, photon self-energies, vertex function renormalization, and the beta function. Section IV presents the results for the two-loop gauge field self-energy. Certain integrals are provided in the appendix, alongside the main text, for the sake of brevity. Lastly, we offer concluding remarks in the final section. It is important to emphasize that throughout this paper, we adopt natural units in which the values of $c$ and $\hbar$ are set to unity, as well as the convention for the spacetime signature of $(+---)$.

\section{Definition of the model}\label{sec11}
  
There are specific $\mathcal{R}\oplus \mathcal{L}$ spin $1/2$ representations of the Lorentz group given in the rest frame by 
\begin{align}
\xi({\bf 0},1)  &= \sqrt{m}\left(
\begin{array}{cc}
0 \\
i \\ 
1 \\
0
\end{array}
\right),\quad
\xi({\bf 0},2) = \sqrt{m}\left(
\begin{array}{cc}
-i \\
0 \\ 
0 \\
1
\end{array}
\right),\label{eq:restlambda}\\
\xi({\bf 0},3) &= \sqrt{m}\left(
\begin{array}{cc}
1\\
0 \\ 
0 \\
-i
\end{array}
\right),\quad
\xi({\bf 0},4) = \sqrt{m}\left(
\begin{array}{cc}
0 \\
1 \\ 
i \\
0
\end{array}
\right),
\end{align}
and  
\begin{align}
\zeta({\bf 0},1) & = \sqrt{m}\left(
\begin{array}{cc}
i \\
0 \\ 
0 \\
1
\end{array}
\right),\quad
\zeta({\bf 0},2) = \sqrt{m}\left(
\begin{array}{cc}
0 \\
i \\ 
-1 \\
0
\end{array}
\right),\\
\zeta({\bf 0},3) &= \sqrt{m}\left(
\begin{array}{cc}
0\\
1\\ 
-i \\
0
\end{array}
\right),\quad
\zeta({\bf 0},4) = \sqrt{m}\left(
\begin{array}{cc}
-1\\
0 \\ 
0 \\
-i
\end{array}
\right),\label{eq:restrho}
\end{align} with the property of being eigenspinors (with eigenvalues $\pm 1$) of the charge conjugation operator
\begin{equation}
\mathcal{C} = 
\left(
\begin{array}{cc}
 \0 & \sigma_2\\
-\sigma_2 &  \0
\end{array}
\right) K,
\end{equation} where $K$ complex conjugates functions and spinors to its right. Therefore, the two above sets of spinors are neutral in this sense. The arbitrary momentum spinors can be derived by applying a boost operator $D(L(p))$ belonging to the direct sum of right-handed and left-handed representations, denoted as $\mathcal{R}\oplus \mathcal{L}$. For further details, please refer to Refs. \cite{Ahluwalia:2022yvk} and \cite{physrep}. Consequently, we define a quantum field whose expansion coefficients are determined by the aforementioned spinors as follows:
\begin{align}
\lambda(x)  = &\int \frac{d^3p}{(2\pi)^3}
\frac{1}{\sqrt{2 m E({\bf p})}}  \nonumber\\
&  \times \sum_\sigma \Big[ a({\bf p},\sigma) \xi({\bf p},\sigma) e^{-i p \cdot x} 
+ b^\dagger({\bf p},\sigma) \zeta({\bf p},\sigma) e^{i p \cdot x} \Big] . \label{eq:field}
\end{align} 

It can be verified that boost and translations are naturally accommodated by the field definition and, in Ref. \cite{Ahluwalia:2022yvk}, it is shown that rotational symmetry is also accomplished as far as the doubling of degrees of freedom, encompassed by the Wigner degeneracy, is present. In this case, the annihilation and creation operators respect fermionic statistics,  leading to the so-called mass dimension one fermionic (MDOF) field. It is important to refer to Refs \cite{Ahluwalia:2022yvk,GNR} for a complete account on the darkness of such a field.    


The dual of Elko spinors cannot be identical to the standard one, as it is evident from the observation that Elkos have zero norms under the Dirac dual. For a formal account on the possibility of dual definitions, one can refer to Appendix A.1 of the Ref. \cite{physrep}. Introducing the quantity $D=\gamma_\mu p^\mu/m$, the Elko dual can be defined by
\begin{align}
\gdualn{\xi}({\bf p},\sigma) & = \left[ + \mathcal{D} \,\xi({\bf p},\sigma) \right]^\dagger \gamma_0, \\
\gdualn{\zeta}({\bf p},\sigma) & =  \left[ - \mathcal{D} \,\zeta({\bf p},\sigma) \right]^\dagger \gamma_0,
\end{align} and hence we are led to 
 \begin{align}
\gdualn{\lambda}(x) = &\int \frac{d^3p}{(2\pi)^3}
\frac{1}{\sqrt{2 m E({\bf p})}}  \nonumber\\
&  \times \sum_\sigma \Big[ a^\dagger({\bf p},\sigma) \gdualn{\xi}({\bf p},\sigma) e^{i p \cdot x} 
+ b({\bf p},\sigma) \gdualn{\zeta}({\bf p},\sigma) e^{-i p \cdot x} \Big],\nonumber
\end{align} as the field adjoint. We can obtain, after applying  \(\lambda\) and\(\gdualn{\lambda}\) upon in or out states, the following Feynman rules for the external legs for the MDOF particle and antiparticle, respectively
\begin{eqnarray*}
 \frac{\xi({\bf p},\sigma)}{\sqrt{m}}
    =  \vcenter{\hbox{\includegraphics[scale=.15]{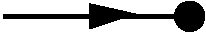}}}
    \qquad\qquad\qquad\frac{\gdualn{\xi}({\bf p},\sigma)}{\sqrt{m}}
    =  \vcenter{\hbox{\includegraphics[scale=.15]{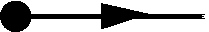}}}\nonumber\\
    \frac{\gdualn{\zeta}({\bf p},\sigma)}{\sqrt{m}}
    = \vcenter{\hbox{\includegraphics[scale=.15]{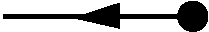}}}\qquad\qquad\qquad
      \frac{\zeta({\bf p},\sigma)}{\sqrt{m}}
    =  \vcenter{\hbox{\includegraphics[scale=.15]{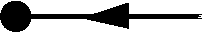}.}} 
\end{eqnarray*}

It can be verified that the Feynman-Dyson propagator for the MDOF field is proportional to the Klein-Gordon one, from which the mass dimension one can be read and the free Hamiltonian is positive definite.   

We now present the Lagrangian that describes the coupling between fermions with mass dimension one ($\lambda$) and a $U(1)$ gauge field, as given in Ref. \cite{Ahluwalia:2022yvk} 
\begin{eqnarray}\label{lag01}
\mathcal{L}=&& -\frac{(1+\delta_3)}{4}F^{\mu\nu}F_{\mu\nu}
+\frac{1}{2}(1+\delta_2)\partial_\mu\gdualn{\lambda} \partial^\mu\lambda -\frac{m^2}{2}(1+\delta_{m^2})\gdualn{\lambda}\lambda \nonumber\\
&& +\tilde{e}(1+\delta_1)\gdualn{\lambda} [\gamma_{\mu},\gamma_\nu]\lambda F^{\mu\nu} -g(1+\delta_g)(\gdualn{\lambda}\lambda)^2+ \mathcal{L}_{GF},
\end{eqnarray}
where $\tilde{e}$ represents the MDOF/gauge-field non-minimal coupling constant and $\mathcal{L}_{GF}$ is the gauge-fixing Faddeev-Popov Lagrangian. The counterterms $\delta_i$ are obtained from field redefinitions $\lambda\rightarrow Z_2^{1/2}\lambda$ and $A^\mu\rightarrow Z_3^{1/2}A^\mu$, along with the following relationships:
\begin{eqnarray}
Z_2&=&1+\delta_2;\nonumber\\
Z_3&=&1+\delta_3;\nonumber\\
m^2 Z_{m^2}&=&(1+\delta_{m^2})m^2=m_0^2Z_2;\nonumber\\
\tilde{e}Z_1&=& \tilde{e}(1+\delta_1)=\tilde{e}_0Z_2 Z_3^{1/2};\nonumber\\
gZ_g &=& g(1+\delta_g)=g_0 Z_2^2,
\end{eqnarray}
where $\tilde{e}_0$ and $g_0$ represent the bare coupling constants. 
 
The propagators of the model in terms of the renormalized quantities are given by
\begin{subequations}
\begin{eqnarray}
S_F(p) &=& \langle T\lambda(p)\gdualn{\lambda}(-p)\rangle= \frac{2i~\bf{1}}{p^2-m^2};\\
\Delta^{\mu\nu}(p) &=& \langle T A^\mu(p)A^\nu(-p)\rangle= -\frac{i}{p^2}\left(\eta^{\mu\nu}-(1-\xi)\frac{p^\mu p^\nu}{p^2} \right).  
\end{eqnarray}
\end{subequations}

\section{The one-loop renormalization}

Let us show the obtained results, case by case, starting from the MDOF self-energy. 

\subsection{The one-loop MDOF self-energy}

\begin{figure}[h!] 	\includegraphics[angle=0 ,width=12cm]{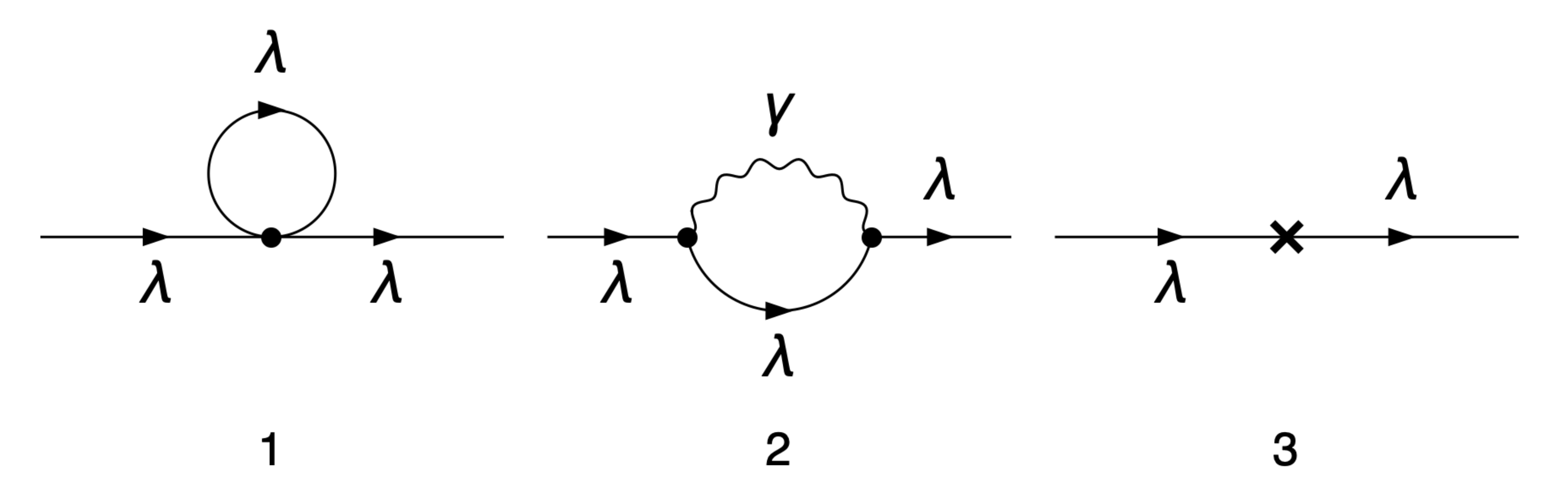} 	\caption{The one-loop ELKO self-energy. Continuous, dashed and wavy lines represent ELKO, auxiliary and photon propagators, respectively. } 	\label{ELKO-SE} \end{figure}

The one-loop contributions to the MDOF self-energy are depicted in Figure \ref{ELKO-SE}. We use a set of MATHEMATICA packages to generate and manipulate the amplitudes\cite{feyncalc,feynarts,feynrules,feynhelpers}.  The corresponding expressions for the first diagram are given by 
\begin{eqnarray}
\Sigma_1(p)= 2ig\int\frac{d^4k}{(2\pi)^4}\frac{\text{tr}{\bf{1}}-1}{k^2-m^2}=-\frac{3g}{8\pi^2} \text{A}_0\left(m^2\right)
=
-\frac{3gm^2}{8\pi^2\epsilon}+\mathrm{finite},
\end{eqnarray}
\noindent where $\epsilon=(4-D)/2$ and $\text{A}_0\left(m^2\right)$ is a Passarino-Veltman (PaVe) integral. We use the same notations and conventions for PaVe integrals as the Ref.\cite{feyncalc}.
 
For the second diagram, we have
\begin{eqnarray}
\Sigma_2(p) &=& 8i \tilde{e} \int\frac{d^4k}{(2\pi)^4}\frac{
\left[(\gamma \cdot (p-k)).\gamma ^{\nu }-\gamma ^{\nu }.(\gamma \cdot (p-k))\right]\left[ (\gamma \cdot (k-p)).\gamma ^{\mu }-\gamma ^{\mu }.(\gamma \cdot (k-p))\right]}{(k^2-m^2)(k-p)^2}\nonumber\\ 
&&\times\left(g^{\mu  \nu }+\frac{\left(1-\xi\right) (k-p)^{\mu } (p-k)^{\nu }}{(k-p)^2}\right)\nonumber\\
&=&\frac{2 \tilde{e}^2 \left[\text{A}_0\left(m^2\right) \left((D-2) (\gamma \cdot p)^2+p^2\right)-(D-2) \left(m^2+p^2\right) \left((\gamma \cdot p)^2-p^2\right) \text{B}_0\left(p^2,0,m^2\right)\right]}{\pi^2 p^2}\nonumber\\ 
&=& \frac{6 \tilde{e}^2 m^2}{\pi^2 \epsilon}
+\mathrm{finite}.
\end{eqnarray} 

Summing these diagrams and adding the counterterm contribution, Figure \ref{ELKO-SE}-(3), we have 
\begin{eqnarray}
\Sigma(p) &=&\frac{1}{2} \left(\delta_2 p^2-m^2\delta_{m^2}\right)-\frac{ 3 m^2 \left(g-16\tilde{e}^2\right)}{8 \pi^2 \epsilon}+\mathrm{finite}.
\end{eqnarray}
Finally, imposing finiteness over minimal subtraction renormalization scheme (MS), we find the counterterms
\begin{eqnarray}\label{eqdelta2}
\delta_{m^2} =\frac{3(16 \tilde{e}^2-g)}{4 \pi^2 \epsilon},\hspace{1cm}
\delta_{2} =0.
\end{eqnarray}
Notice that, unlike the Dirac fermions coupled to photons, the wave-function counterterm is vanishing for the MDOF field at one-loop order. Besides, Ref. \cite{deBrito:2019hih} demonstrated that the MDOF four-point function is finite, and this property persists even in gauge interactions.

\subsection{The photon self-energy}

\begin{figure}[h!] 	\includegraphics[angle=0 ,width=8cm]{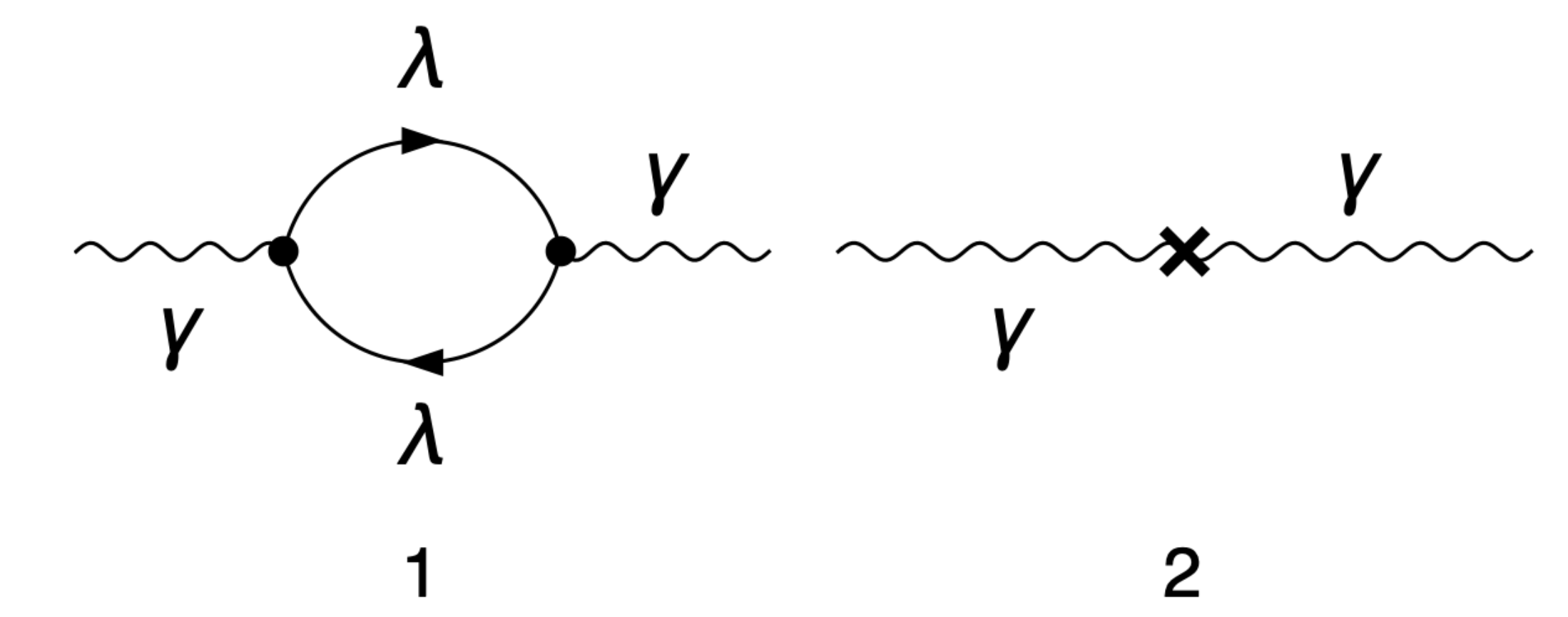} 	\caption{The one-loop photon self-energy.} 	\label{Photon-SE} \end{figure}

The diagrams of the photon self-energy are depicted in Fig. \ref{Photon-SE}. The corresponding expression is given by
\begin{eqnarray}\label{photon-se}
\Pi^{\mu\nu}(p) &=& -8i\tilde{e}^2~\text{tr}\int\frac{d^4k}{(2\pi)^4}\frac{ \left[(\gamma \cdot p)\gamma^{\nu}-\gamma^{\nu}(\gamma \cdot p)\right] \left[ (\gamma \cdot p).\gamma ^{\mu }- \gamma^{\mu } (\gamma \cdot p)\right]}{\left(k^2-m^2\right)\left((k-p)^2-m^2\right)}\nonumber\\
&=& -\frac{16 \tilde{e}^2}{\pi^2} \text{B}_0\left(p^2,m^2,m^2\right)\left(p^2 g^{\mu  \nu }-p^{\mu } p^{\nu }\right) \nonumber\\
&=&\frac{16 \tilde{e}^2}{\pi^2\epsilon} \left(p^2 g^{\mu  \nu }-p^{\mu } p^{\nu }\right)+\mathrm{finite}. 
\end{eqnarray}

Adding the couterterm and imposing finiteness, we find
\begin{eqnarray}\label{eqdelta3}
\delta_{3} =\frac{16\tilde{e}^2}{\pi^2 \epsilon}.
\end{eqnarray}
This is a remarkable result. In Ref. \cite{Ahluwalia:2004ab}, it was mentioned by the authors that the interaction term $\tilde{e}\bar\lambda [\gamma_{\mu},\gamma_\nu]\lambda F^{\mu\nu}$ could impact the propagation of the photon by inducing an effective mass term. However, as indicated by Eq.\eqref{photon-se}, it appears that this is not the case, as the one-loop photon propagation remains transverse and massless. We shall revisit this point in the next section, studying two-loop effects.

\subsection{The vertex function renormalization}

\begin{figure}[h!] 	\includegraphics[angle=0 ,width=10cm]{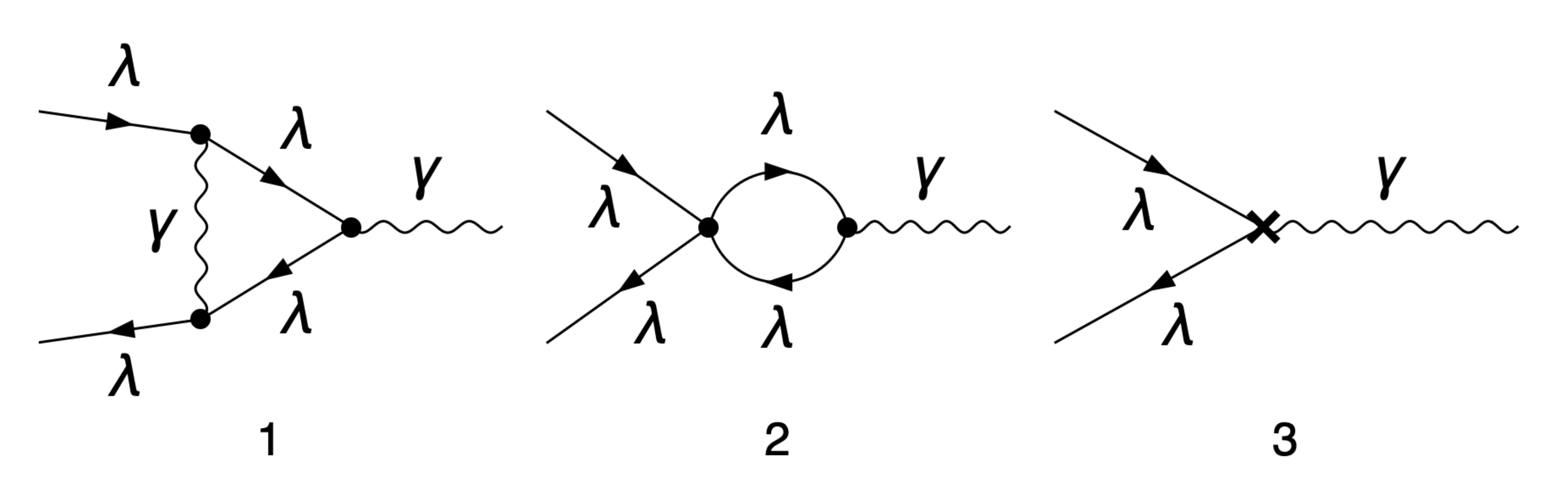} 	\caption{The one-loop corrections to the vertex function.} 	\label{vertex1} \end{figure}

The diagrams of the electromagnetic vertex function are depicted in Fig. \ref{vertex1}. The corresponding expression is given by
\begin{eqnarray}
\Gamma^{\mu}_1(p_1,p_2) &=& -128\tilde{e}^3\int\frac{d^4k}{(2\pi)^4}\frac{\left(\gamma ^{\mu }~\gamma \cdot (p_1+p_2)
-\gamma \cdot (p_1+p_2) \gamma ^{\mu}\right)}{\left((k+p_2)^2-m^2\right)\left((k-p_1)^2-m^2\right)}\nonumber\\
&=& -\frac{8i\tilde{e}^3}{\pi^2}\left[\gamma ^{\mu }~\gamma \cdot (p_1+p_2)
-\gamma \cdot (p_1+p_2) \gamma ^{\mu}\right] \text{B}_0\left((p_1+p_2)^2,m^2,m^2\right) \nonumber\\
&=&- \frac{8i\tilde{e}^3}{\pi^2\epsilon}\left[\gamma ^{\mu }~\gamma \cdot (p_1+p_2)
-\gamma \cdot (p_1+p_2) \gamma ^{\mu}\right] +\mathrm{finite},
\end{eqnarray}
\noindent where we have used the momentum conservation $p_3=p_1+p_2$.

The second diagram gives
\begin{eqnarray}
\Gamma^{\mu}_2(p_1,p_2) &=& 4\tilde{e}g\int\frac{d^4k}{(2\pi)^4}\frac{\left(\gamma ^{\mu }~\gamma \cdot (p_1+p_2)
-\gamma \cdot (p_1+p_2) \gamma ^{\mu}\right)}{\left((k+p_2)^2-m^2\right)\left((k-p_1)^2-m^2\right)}\nonumber\\
&=& \frac{i\tilde{e}g}{2\pi^2}\left[\gamma ^{\mu }~\gamma \cdot (p_1+p_2)
-\gamma \cdot (p_1+p_2) \gamma ^{\mu}\right] \text{B}_0\left((p_1+p_2)^2,m^2,m^2\right) \nonumber\\
&=& \frac{i\tilde{e}g}{2\pi^2\epsilon}\left[\gamma ^{\mu }~\gamma \cdot (p_1+p_2)
-\gamma \cdot (p_1+p_2) \gamma ^{\mu}\right] +\mathrm{finite}.
\end{eqnarray}

Adding these contributions to the vertex counterterm, we have
\begin{eqnarray}
\Gamma^{\mu}(p_1,p_2) &=& -i\left(\frac{\tilde{e}(16\tilde{e}^2-g)}{2\pi^2\epsilon}+2\tilde{e}\delta_1 \right) \left[\gamma ^{\mu }~\gamma \cdot (p_1+p_2)
-\gamma \cdot (p_1+p_2) \gamma ^{\mu}\right] +\mathrm{finite},
\end{eqnarray}
 
\noindent from which imposing finiteness we find
\begin{eqnarray}\label{eqdelta1}
\delta_1 &=&\frac{g-16\tilde{e}^2}{4\pi^2\epsilon}.
\end{eqnarray}

\subsection{The one-loop beta function}

The Lagrangian given in Eq. \eqref{lag01} shows that the relation between bare and renormalized couplings is expressed as $Z_1\tilde{e}=\mu^{-2\epsilon}\tilde{e}_0Z_2Z_3^{1/2}$. The MDOF wave-function renormalization constant $Z_2$, equal to $(1+\delta_2)$, is unity. As a result, the beta function for the electric charge can be expressed as
\begin{eqnarray}
\beta(\tilde{e})=\lim_{\epsilon\rightarrow 0}\mu\frac{d\tilde{e}}{d\mu}=\frac{\tilde{e}}{2\pi^2}\left(g-48\tilde{e}^2\right),
\end{eqnarray}
\noindent where we can see that the non-minimal electromagnetic coupling can be asymptotically free if $g<48\tilde{e}^2$. 

\section{The two-loop gauge field self-energy}

\begin{figure}[h!] 	\includegraphics[angle=0 ,width=10cm]{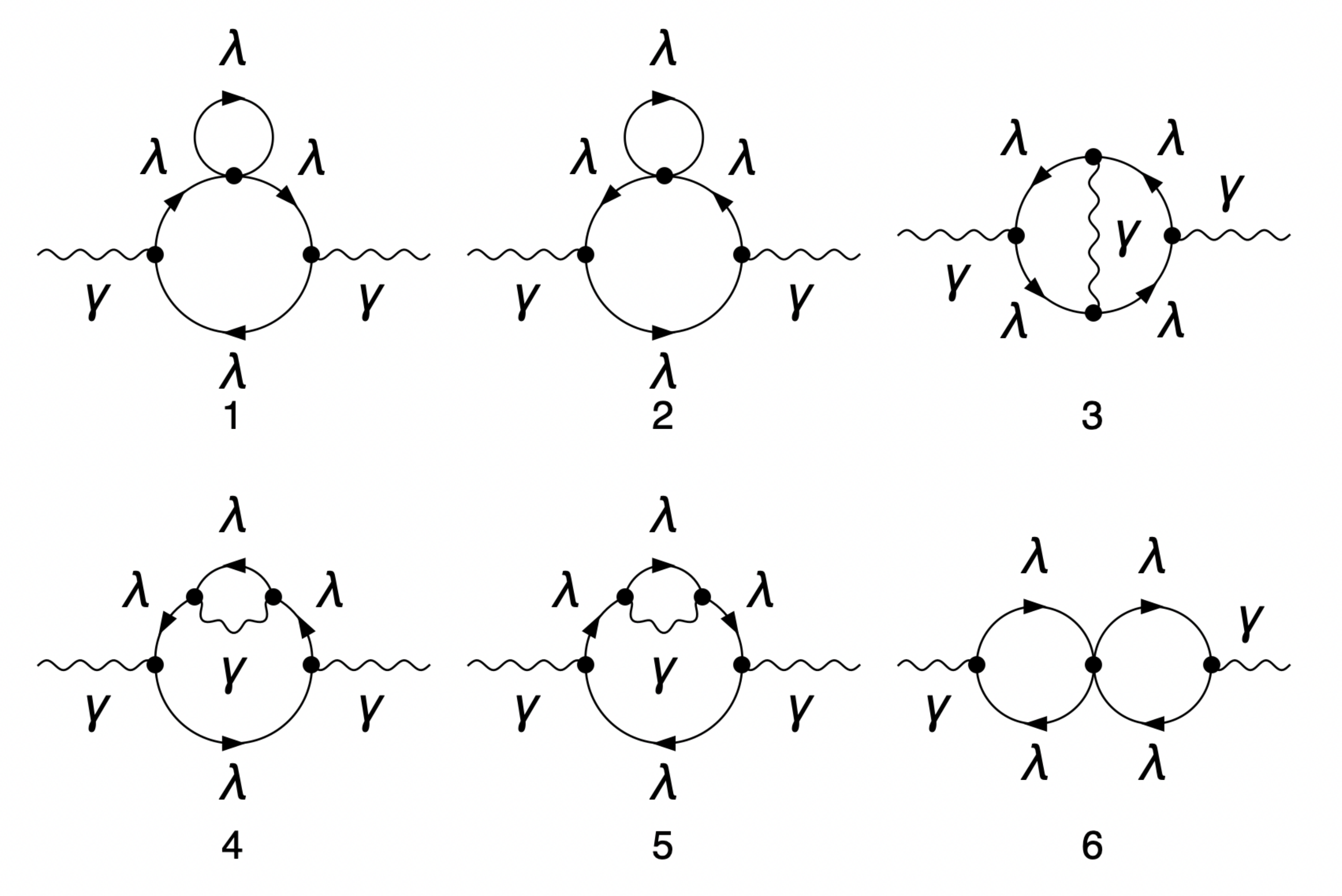} 	\caption{The two-loop corrections to the gauge field propagation.} 	\label{gauge2l} \end{figure}

\begin{figure}[h!] 	\includegraphics[angle=0 ,width=12cm]{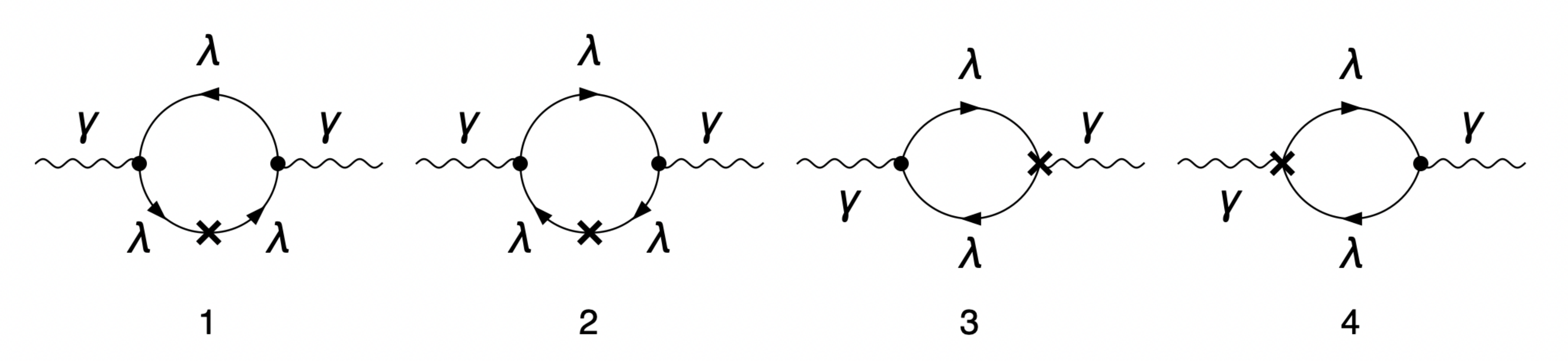} 	\caption{One-loop corrections to the propagation of the gauge field with the insertion of a single counterterm.} 	\label{gauge2lCT} \end{figure}

This section focuses on obtaining the two-loop corrections to the gauge-field propagation. This process involves the two-loop diagrams depicted in Fig.\ref{gauge2l} and the one-loop diagrams with a counterterm insertion shown in Fig.\ref{gauge2lCT}.

Our approach to computing the two-loop diagrams is similar to that described in Ref.\cite{Bevilaqua:2021uzk}. Naturally, each diagram $i$ should have the most general (Lorentz Invariant) form $\Pi_i^{\mu\nu}(p)=\eta^{\mu\nu}p^2~A_i(p)+p^\mu p^\nu B_i(p)$, from  which $A_i(p)$ and $B_i(p)$ can be obtained through the projections:
\begin{eqnarray}
A_i &=& \frac{1}{(D-1)p^2}\left(\eta_{\mu\nu}-\frac{p_{\mu}p_\nu}{p^2}\right)\Pi_i^{\mu\nu},\nonumber\\
B_i &=& -\frac{1}{(D-1)p^2}\left(\eta_{\mu\nu}-D\frac{p_{\mu}p_\nu}{p^2}\right)\Pi_i^{\mu\nu}. \nonumber
\end{eqnarray}
By summing over the diagrams, we obtain the expected result $\Pi(p)=\sum_i A_i(p)=-\sum_i B_i(p)$, which implies that the photon polarization tensor has the transverse form. In these calculations, we reduced the scalar two-loop integrals using the Tarasov algorithm\cite{Tarasov:1997kx}, with the assistance of the computational package TARCER~\cite{Mertig:1998vk}. The resulting scalar two-loop integrals table can be found in \cite{Martin:2003qz}.

The two-loop gauge field self-energy, depicted in Fig.\ref{gauge2l}, is given by
\begin{eqnarray}
\Pi^{\mu\nu}_{2l}(p)=(\eta^{\mu\nu}p^2-p^\mu p^\nu)\Pi_{2l}(p),
\end{eqnarray}
\noindent where
\begin{eqnarray}
\Pi_{2l}(p) &=&\frac{1024 (D-2) \tilde{e}^2~(\text{A}_0(m^2))^2}{(4\pi)^D(D-1) m^2 p^2 \left(4 m^2-p^2\right)}\left[16 \tilde{e}^2 \left(\left(-D^2+D+3\right) p^2+4 (D-4) m^2\right)+3 (D-1) g p^2\right]\nonumber\\
&&+\frac{16384 \tilde{e}^2~\text{A}_0(m^2)\text{B}_0(p^2,m^2,m^2)}{(4\pi)^D (D-1) m^2 \left(4 m^2-p^2\right)} \Big[\tilde{e}^2 \left(2 (D ((D-7) D+19)-19) m^2+(D-4) (D-2) p^2\right)\nonumber\\
&& -\frac{3}{8} (D-3) (D-1) g m^2\Big]+\frac{1024 \tilde{e}^2 (\text{B}_0(p^2,m^2,m^2))^2}{(4\pi)^D} \left[\frac{16 ((D-8) D+13) \tilde{e}^2}{D-1}+g\right]\nonumber\\
&&- \frac{32768 (D-4) \tilde{e}^4}{(4\pi)^D (D-1) p^2}\left[(D-3) \text{J}^{(D)}_{\{1,m\},\{1,m\},\{1,0\}} + \left(p^2-4 m^2\right) \text{J}^{(D)}_{\{2,m\},\{1,m\},\{1,0\}} \right],
\end{eqnarray}
\noindent with the integrals $J$ defined as in Ref.\cite{Mertig:1998vk}. Expanding above expression around $D=4$ and $p^2\approx 0$, we have
\begin{eqnarray}\label{eq2l}
\Pi_{2l}(p) = -\frac{4 \tilde{e}^2 \left(16 \tilde{e}^2-g\right)}{\pi^4 \epsilon^2}+\frac{4 \tilde{e}^2 \left(16 \tilde{e}^2-g\right)}{\pi^4 \epsilon} \left(2 \log\left(\frac{m^2}{\mu^2}\right) +2 \gamma_E -3-2 \log{(4\pi)} \right) +\mathcal{O}(p^2),
\end{eqnarray}
\noindent where $\epsilon=(4-D)/2$ and $\mu$ is a mass scale introduced by the dimensional regularization.

The one-loop diagrams with insertions of counterterms are given by
\begin{eqnarray}\label{eq2lct}
\Pi_{CT}(p) &=& -\frac{32 \tilde{e}^2}{\pi^2}\left[ (\delta_1-\delta_2) \text{B}_0\left(p^2,m^2,m^2\right)+ m^2 (\delta_{m^2}-\delta_2) \text{C}_0\left(0,p^2,p^2,m^2,m^2,m^2\right)\right]\nonumber\\
&=& \frac{16\tilde{e}^2}{\pi^2}\left[(4\delta_1-5\delta_2+\delta_{m^2})+2\frac{(\delta_1-\delta_2) (\epsilon  (\gamma_E -2-\log(4 \pi))-1)}{\epsilon } +2(\delta_1-\delta_2)\log\left(\frac{m^2}{\mu^2}\right)\right]\nonumber\\
&&+\mathcal{O}(p^2).
\end{eqnarray}
Substituting the counterterms Eqs. \eqref{eqdelta2} and \eqref{eqdelta1} into \eqref{eq2lct}, and adding it to the Eq. \eqref{eq2l}, we finally have the UV divergent part of the two-loop corrections to the gauge field propagation as
\begin{eqnarray}
\Pi^{\mu\nu}_{2l}(p^2)=\left(\frac{4 \tilde{e}^2 \left(16 \tilde{e}^2-g \right)}{\pi ^4\epsilon^2}+\mathcal{O}(p^2)\right) \left(\eta^{\mu\nu}p^2-p^\mu p^\nu \right).
\end{eqnarray}

It is important to emphasize that the interaction term $\tilde{e}\gdualn{\lambda} [\gamma_{\mu},\gamma_\nu]\lambda F^{\mu\nu}$ does not lead to an effective mass term, contrary to what is suggested in \cite{Ahluwalia:2004ab}, as it can be seen from Eqs. \eqref{eq2l} and \eqref{eq2lct}, the photon propagation remains massless and transverse up to the two-loop order.
 
\section{Final Remarks}\label{summary}

Our study focuses on investigating the role of the MDOF field in Quantum Electrodynamics (QED) within the framework of renormalization at one and two-loop orders. Notably, we observe a distinction between Dirac and MDOF field regarding the vanishing wave-function counterterm. Additionally, our results reveal that the photon propagation remains massless and transverse up to the two-loop order. An important result of our study is that the non-minimal electromagnetic coupling can exhibit asymptotic freedom if the condition $g<48\tilde{e}^2$ is satisfied.

This finding has profound implications for the behavior of the MDOF field at high energies. It suggests that the MDOF field has the potential to decouple from other particles and interactions at these energy scales. This aspect is relevant when considering the MDOF field as a candidate for dark matter. It implies that the MDOF field may behave differently from other matter fields at high energies, which has implications for calculating its relic abundance and freeze-out temperature. In particular, it impacts processes involving the interactions between the MDOF field and the ordinary particles of the Standard Model.

Furthermore, future investigations can expand upon our calculations by incorporating gravitational corrections and exploring higher energy scales. Such analyses would shed light on more intriguing aspects of the primordial universe and cosmological applications involving the MDOF particle as a constituent of dark matter.

\acknowledgments
We would like to thank D.~V.~Ahluwalia for useful discussion and comments. W.C. is partially supported by Coordena\c{c}\~ao de Aperfei\c{c}oamento de Pessoal de N\'ivel Superior (CAPES). J.M.H.S. thanks CNPq (grant No. 307641/2022-8) for financial support. 

\appendix
\section{Integrals notations}

We are adopting the same notations and conventions for Passarino-Veltman (PaVe) one-loop integrals as the Ref.\cite{feyncalc}. The integrals appearing in our work are
\begin{eqnarray}
\text{A}_0(m^2)&=& \int\frac{d^Dk}{i\pi^2}\frac{1}{k^2-m^2};\\
\text{B}_0(p^2,m^2,m^2)&=& \int\frac{d^Dk}{i\pi^2}\frac{1}{(k^2-m^2)((k-p)^2-m^2)};\\
\text{C}_0\left(0,p^2,p^2,m^2,m^2,m^2\right)&=& \int\frac{d^Dk}{i\pi^2}\frac{1}{(k^2-m^2)^2((k+p)^2-m^2)}.
\end{eqnarray}

The two-loop integrals in this work are labeled in accordance with Ref. \cite{Mertig:1998vk}. We will now provide a quotation of the two-loop integrals that appear in our calculations:
\begin{eqnarray}
\text{J}^{(D)}_{\{1,m\},\{1,m\},\{1,0\}}&=&\iint\frac{d^Dk_1~d^Dk_2}{\pi^D}\frac{1}{(k_1^2-m^2)((k_1-k_2)^2-m^2)(k_2-p)^2}; \\
\text{J}^{(D)}_{\{2,m\},\{1,m\},\{1,0\}} &=& \iint\frac{d^Dk_1~d^Dk_2}{\pi^D}\frac{1}{(k_1^2-m^2)^2((k_1-k_2)^2-m^2)(k_2-p)^2}.
\end{eqnarray}

\newpage

\end{document}